\documentclass[%
 aip,
 jmp,%
 amsmath,amssymb,
reprint,%
twocolumn]{revtex4-1}

\usepackage{dcolumn}% Align table columns on decimal point
\usepackage{bm}% bold math

\usepackage{latexsym}        %packages required for symbols in revtex4.
\usepackage{amssymb}
\usepackage{amsmath}
\usepackage{amsfonts}
\usepackage[dvips]{graphicx}

\begin{document}

%\preprint{APS/123-QED}

\title{Compressed correlation functions and fast aging dynamics in metallic glasses}% Force line breaks with \\

\author{B. Ruta}%
 \email{ruta@esrf.fr}
\affiliation{European Synchrotron Radiation Facility, BP220, F-38043 Grenoble, France.}
\author{G. Baldi}%
\affiliation{IMEM-CNR Institute, Parco Area delle Scienze, I-43124 Parma, Italy}
\author{G. Monaco}%
\affiliation{European Synchrotron Radiation Facility, BP220, F-38043 Grenoble, France.}
\author{Y. Chushkin}%
\affiliation{European Synchrotron Radiation Facility, BP220, F-38043 Grenoble, France.}
%\author{L. Cipelletti}%
%\affiliation{Universit\'{e} Montpellier 2, Laboratoire Charles Coulomb UMR 5221, F-34095, Montpellier, France.}
%\author{E. Pineda}%
%\affiliation{Departament de F\'{i}sica i Enginyeria Nuclear, ESAB, Universitat Polit\'{e}cnica de Catalunya, c/ Esteve Terradas %8, 08860 Castelldefels, Spain.}
%\author{P. Bruna}
%\affiliation{Departament de F\'{i}sica Aplicada, EPSC, Universitat Polit\'{e}cnica de Catalunya, c/ Esteve Terradas 5, 08860 %Castelldefels, Spain.}
\textbackslash\textbackslash
\date{\today}% It is always \today, today,

\begin{abstract}
We present x-ray photon correlation spectroscopy measurements of the atomic dynamics in a Zr$_{67}$Ni$_{33}$ metallic glass, well below its glass transition temperature. We find that the decay of the density fluctuations can be well described by compressed, thus faster than exponential, correlation functions which can be modeled by the well-known Kohlrausch-Williams-Watts function with a shape exponent $\beta$ larger than one. This parameter is furthermore found to be independent of both waiting time and wave-vector, leading to the possibility to rescale all the correlation functions to a single master curve. The dynamics in the glassy state is additionally characterized by different aging regimes which persist in the deep glassy state. These features seem to be universal in metallic glasses and suggest a non diffusive nature of the dynamics. This universality is supported by the possibility of describing the fast increase of the structural relaxation time with waiting time using a unique model function, independently of the microscopic details of the system. 
\end{abstract}

\pacs{64.70.pe,65.60.+a,64.70.pv}% PACS, the Physics and Astronomy
\maketitle

\section{\label{sec:level1}Introduction}

Understanding physical aging in glasses is essential at a technological level, for controlling the material properties and their temporal evolution, as well as at a fundamental level, since glasses are often considered as archetypes for systems far from thermodynamical equilibrium \cite{angell2000b,berthier2011}. However, and despite the strong efforts done in the field, a complete picture of the dynamics in the glassy state is still missing. The main difficulty arises from the complexity of the glassy properties. When a supercooled liquid is cooled below its glass transition temperature, $T_g$, the system falls out of equilibrium in a metastable state which depends on the previous thermal history, and which evolves, with waiting time, through different states toward the corresponding supercooled equilibrium liquid phase. The existence of a multitude of different metastable glassy states which can be explored through different annealing procedures or cooling rates, makes difficult the development of a universal theory for glasses or even a plain comparison between different experimental results.\\
Physical aging in polymeric, metallic or more conventional inorganic glasses, is commonly studied by looking at the temporal evolution of a given observable, keeping fixed other external parameters, such as the temperature and the pressure of the system. Macroscopic quantities like viscosity, refractive index, volume or elastic constants, are usually the subject of this kind of studies \cite{struik1978,busch1998,napolitano1968,kovacs1958,wen2006,miller1997,grigera1999,boucher2011}. 
In these works, the investigated quantities evolve with waiting time, $t_w$, in a way which depends on the previous thermal history and which can be accurately described by stretched exponential functions \cite{kovacs1958}. Similar results have been reported also in studies of aging effect on the intensity of the high frequency tail of the dielectric spectra \cite{schlosser1991,alegria1997,leheny1998,lunkenheimer2005,lunkenheimer2006,richert2010}. All these works provide information on the equilibration toward the supercooled liquid phase, but they do not give any microscopic details on the aging dynamics in the glassy state. This information can instead be obtained by the investigation, on a microscopic scale, of the temporal evolution of the structural relaxation time, $\tau$, which represents the time necessary for the system to rearrange its structure toward a more stable configuration. Its value can be measured, for instance, by monitoring the long time decay of the intermediate scattering function, $f(q,t)$, on a length scale $2\pi/q$, with the wave-vector $q$ corresponding to the mean inter particle distance. The intermediate scattering function is related to the density-density correlation function and contains all the information on the relaxation dynamics in the system on a length scale determined by $q$. While the dynamics in the supercooled liquid phase has been widely investigated in the past, the difficulty of probing the particle-level dynamics in realistic molecular glasses both with experiments and simulations has instead strongly delayed progress in the field. In fact only very few studies exist of the behavior of $\tau$ below $T_g$ \cite{struik1978,alegria1995,alegria1997,kob1997,kob2000,masri2010,prevosto2004,casalini2009,ruta2012,leitner2012}. Numerical simulations on a Lennard Jones glass show that for small quenching from an equilibrium configuration, $\tau$ slows down linearly or sub linearly with waiting time \cite{kob1997} in agreement with previous results on polymeric glasses \cite{struik1978,alegria1995,alegria1997}. In addition, the long time decay of the correlation functions can be rescaled into a single master curve, leading to the validity of a time-waiting time superposition principle in the glassy state \cite{kob1997,hecksher2010}.\\
Recently, the aging of a Mg$_{65}$Cu$_{25}$Y$_{10}$ metallic glass has been studied in both the supercooled liquid phase and the glassy state by looking directly at the evolution of the dynamics at the interatomic length scale \cite{ruta2012}. The glass transition has been found to be accompanied by i) a crossover of the intermediate scattering function from the well characterized stretched exponential shape - thus described by a stretching exponent $\beta<1$ - in the supercooled liquid phase, to an unusual compressed exponential form - defined by a parameter $\beta>1$ - in the glassy state and ii) to the presence of two very distinct aging regimes below $T_g$. In particular, a fast, exponential growth of $\tau$ has been reported for short waiting times, followed by an extremely slow aging regime at larger $t_w$. While the latter slow regime has been already reported both in numerical simulations \cite{kob1997} and in experiments on polymeric glasses \cite{struik1978,alegria1995,alegria1997}, the compressed shape of the correlation functions and the fast aging regime had not been reported before in structural glasses. Two similar aging regimes have been recently observed in a numerical study of a concentrated system of nearly hard spheres \cite{masri2010}. In this case, however, the correlation functions display a more stretched exponential behavior, and no signature of a compressed shape was reported \cite{masri2010}. \\
Compressed correlation functions and different aging regimes have instead been reported for out of equilibrium soft materials, where the dynamics is controlled more by ballistic-like rather than by diffusive motions \cite{bouchaud2001,cipelletti2003}. The previous work on the metallic glass thus suggests a common nature of the dynamics in metallic glasses and soft materials. A very similar faster than exponential decay and strong aging behavior for the intermediate scattering function has been observed also in a Zr-based metallic glass by Leitner and coworkers \cite{leitner2012}. In this case, however, this unusual dynamics has been associated with the onset of a crystallization process.
There are consequently several questions to clarify: i) are the results found in Mg$_{65}$Cu$_{25}$Y$_{10}$ really a universal feature of metallic glasses or do they depend on the microscopic details of the system? ii) is it possible to find a common description of physical aging for different glasses? iii) what is the effect played by the wave-vector, $q$, on the dynamics?\\
Here, we answer to these questions by presenting a detailed investigation on physical aging in a Zr$_{67}$Ni$_{33}$ metallic glass at the atomic length scale. We find that in this case the dynamics in the glassy state is also characterized by compressed correlation functions and different aging regimes, supporting the idea of a universal mechanism for physical aging in metallic glasses. This hypothesis is moreover strengthened by the possibility of describing the aging dynamics using the same phenomenological exponential growth and rescaling the results obtained for different systems into a single master curve.\\

\section{\label{sec:level2}Experimental details}

The Zr$_{67}$Ni$_{33}$ samples ($T_g=647$ K \cite{georgarakis2010}) were produced as thin ribbons by melt-spinning at the Polytechnic University of Catalonia. The Zr and Ni pure metals were pre-alloyed by arc-melting under Ti-gettered Ar atmosphere. The melt was then fast-quenched with a cooling rate of $10^6$ K/s by injecting it on a Cu wheel spinning with a 40m/s perimeter velocity. The resulting metallic ribbons have a thickness of 22$\pm$4 $\mu$m, close to the optimal value required to maximize the scattered intensity in the experiments. The ribbons were then held in a resistively heated furnace covering a temperature range between 270 K and 1000 K with a thermal stability of less than 1 K. The temperature of the sample was held at $T=373$ K, reached by heating from room temperature with a fixed rate of $3$ K/min.\\
X-ray photon correlation spectroscopy (XPCS) measurements were carried out at the beamline ID10 at the European Synchrotron Radiation Facility in Grenoble, France. An incident x-ray beam of 8.0 keV ($\lambda=1.55$ \r{A}) was selected by using a single bounce Si(1,1,1) crystal monochromator with a bandwidth of $\Delta\lambda/\lambda\approx 10^{-4}$. Hard X-rays originating from higher order monochromator reflections were suppressed by a Si mirror placed downstream of the monochromator. A partially coherent beam with $\sim10^{10}$ photons/s was then obtained by rollerblade slits opened to $10\times10$ $\mu$m, placed $\sim$0.18 m upstream of the sample.
In order to enhance the scattered intensity, speckle patterns were collected in transmission geometry by two IkonM charge-coupled devices from Andor Technology ($1024\times1024$ pixels, $13\times13$ $\mu m$ pixel size each one) installed in the vertical scattering plane $\sim$70 cm downstream of the sample. All pixels of the two CCD detectors were considered to belong to the same $q$ with a resolution of $\Delta q= 0.04$ \r{A}$^{-1}$. The detectors were mounted on a diffractometer and could rotate around the center of the sample position, covering a $q$ range up to several \r{A}$^{-1}$. XPCS measurements were performed for different wave-vectors at and around the first diffraction peak of the static structure factor which corresponds to $q_p=2.55$ \r{A}$^{-1}$, thereby probing directly the temporal relaxation of density fluctuations at the inter-particle distance $2\pi/q_p\sim2$ \r{A}. Time series of up to $\sim$5000 images were taken with 7 s exposure time per frame and were analyzed following the procedure described in Ref. \cite{chushkin2012}. No beam damage was observed and the amorphous structure of the sample was checked during the whole experiment by measuring the corresponding static structure factor of the system.

\section{Results and Discussion}

As previously discussed, physical aging implies a relaxation dynamics which is not the same at all times, but strongly depends on the sample age or waiting time. The temporal dependence of the dynamics can be entirely captured by XPCS through the determination of the two-time correlation function $G(t_1,t_2)$ (TTCF) which represents the instantaneous correlation of the intensity $I$ at two times $t_1$ and $t_2$, being:
\begin{equation}
G(t_1,t_2)=\frac{\left\langle I(t_1)I(t_2)\right\rangle_p}{\left\langle I(t_1)\right\rangle_p\left\langle I(t_2)\right\rangle_p}.
\end{equation}
Here, the average is performed on all the pixels of the CCD, that correspond to the same wave-vector $q$ \cite{chushkin2012,sutton2003,madsen2010}.\\ 
%%%%%%%%%%%%%%%%%%%%%%%%%%%%%%%%%%%%%%%%%%%%%%%%%%%%%%%%%%%%%%%%%%%%%%%%
\begin{figure}
\begin{tabular}{c}
        \centerline{\includegraphics[width=9 cm]{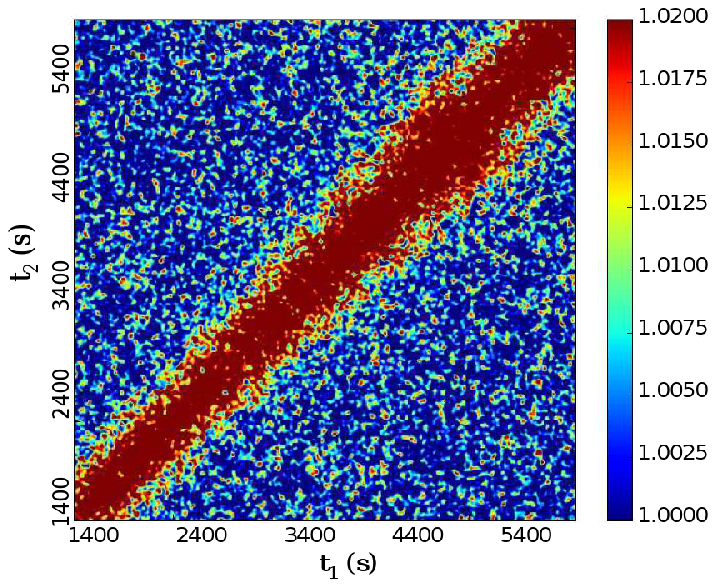}}\\
        \centerline{\includegraphics[width=9 cm]{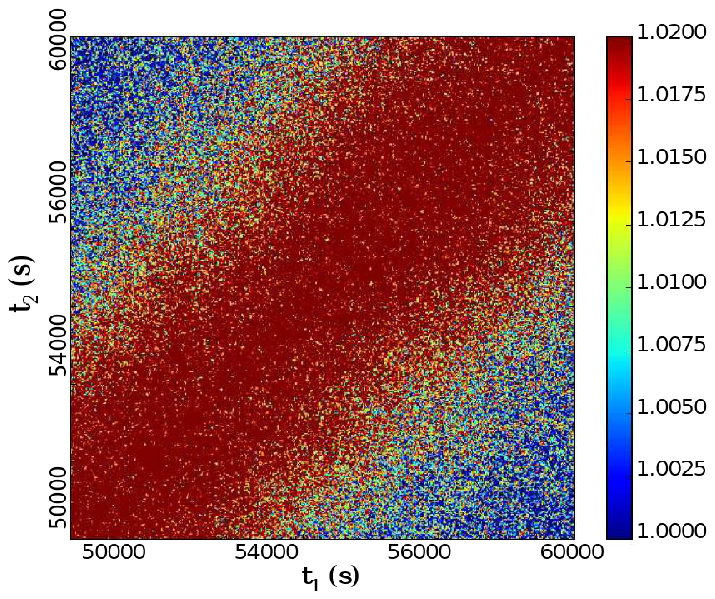}}
        \end{tabular}
\caption{Two-time correlation functions measured with XPCS in Zr$_{67}$Ni$_{33}$ at $T$=373 K. Data acquisition started at $t_0$=1400 s (top panel) and $t_0$=49000 s (bottom panel) from temperature equilibration. The age of the sample at any point in the Figure can be taken as $t_{age}=t_0+(t_1+t_2)/2$ s. The huge slow down of the dynamics at the different sample ages is well illustrated by the clear broadening of the intensity around the diagonal from the left bottom corner to the top right corner in the bottom panel with respect to the case corresponding to a much shorter $t_w$ (top panel).} \label{Fig1}
\end{figure}
%%%%%%%%%%%%%%%%%%%%%%%%%%%%%%%%%%%%%%%%%%%%%%%%%%%%%%%%%%%%%%%%%%%%%%%%
Figure \ref{Fig1} shows the TTCF measured in Zr$_{67}$Ni$_{33}$ in the glassy state at two different sample age times from temperature equilibration. The colors represent the values of $G$% in the point ($t_1,t_2)$, with the darkest being the largest ($G\sim1.02$)
. The diagonal from bottom left to top right corresponds to the time line of the experiment and it thus indicates the sample age. The width of this diagonal contour is instead proportional to the relaxation time $\tau$. 
The steady slowing down of the dynamics, and thus the increase in the relaxation time due to physical aging, is illustrated by the broadening of the diagonal contour on increasing the sample age. This effect is clearly visible in the top panel of Figure \ref{Fig1} where we report data taken between $\sim 1000$ and $\sim5000$ s after temperature equilibration. After almost 13 hours of annealing (lower panel of Figure \ref{Fig1}), the dynamics was so slow that the intensity along the diagonal covered almost the whole image, despite the time interval was almost the double with respect to the one reported in the top panel. After this very long annealing the width of the diagonal contour seems to be constant with time, as it would be in the case of stationary equilibrium dynamics. As discussed later, here the effect of physical aging is probably so slow that the dynamics appears as stationary at least on our experimental time scale.\\ 
Although the dynamics is not stationary, the aging is sufficiently slow that it is possible to find time intervals over which the $G(t_1,t_2)$ can be averaged and thus to get the standard intensity autocorrelation function 
\begin{equation}
g_2(q,t)=\frac{\left\langle \left\langle I(q,t_1)I(q,t_1+t)\right\rangle_p\right\rangle}{\left\langle\left\langle I(q,t_1)\right\rangle_p\left\langle I(q,t_1)\right\rangle_p\right\rangle}
\label{g2}
\end{equation}
where $\left\langle ...\right\rangle$ is the temporal average over all the possible starting times $t_1$. 
The intensity autocorrelation function is related to the intermediate scattering function through the Siegert relation $g_2(q,t)=1+B(q)\left|f(q,t)\right|^2$, with $B(q)$ a setup-dependent parameter \cite{madsen2010}, and its determination allows capturing snapshots of the dynamics at different waiting times. The Siegert relation requires the temporal average to be equal to the ensemble average. This condition is fulfilled for systems at thermodynamic equilibrium. The extension to non-ergodic systems, such as glasses, is possible thanks to the use of an area detector, like the CDD used during our experiments. In this case in fact, the intensity distribution is sampled correctly by the average over all the pixels, independently of the nature of the dynamics \cite{cipelletti1999,bartsch1997}.\\
The time interval for temporal averaging was chosen small enough that the shape of $g_2(q,t)$ was not noticeably affected by the evolution of the dynamics, while keeping a good signal to noise ratio. For each set of images, the sample age was defined as $t_w=t_0+(t_f-t_i)/2$, where $t_0$ is the delay of the starting point for the measurement from temperature equilibration, and $t_f$ and $t_i$ are the final and starting times of the interval used for the average.\\
The $\alpha$ relaxation process in glass-formers can be well described by using the empirical Kohlrausch Williams Watt (KWW) model function \cite{angell2000b}
\begin{equation}
f(q,t)=f_q(T)\cdot \exp[-(t/\tau)^\beta]
\end{equation}
In this expression, $\tau$ is the characteristic relaxation time, $\beta$ is the stretching
parameter which quantifies the deviation from a simple exponential behavior, and $f_q(T)$ is the nonergodicity factor. It represents the height of the intermediate-time plateau before the final decay
associated to the $\alpha$-relaxation and it slightly depends on temperature in the glassy state.\\ 
Following the Siegert relation, the intensity correlation functions measured with XPCS can then be described by
\begin{equation}
g_2(q,t)=1+c(q,T)\cdot \exp[-2(t/\tau)^\beta]
\label{KWW}
\end{equation} 
where $c=B(q) f_q^2$, being $B(q)$ the parameter entering the Siegert relation.\\
A selection of normalized intensity autocorrelation functions obtained at $T=373$ K for a fixed $q=q_p$ and different waiting times is reported in Figure \ref{Fig2}. The data are shown together with the best-fit line shapes obtained by using equation \eqref{KWW}.
%%%%%%%%%%%%%%%%%%%%%%%%%%%%%%%%%%%%%%%%%%%%%%%%%%%%%%%%%%%%%%%%%%%%%%%%
\begin{figure}
\includegraphics[width=9 cm]{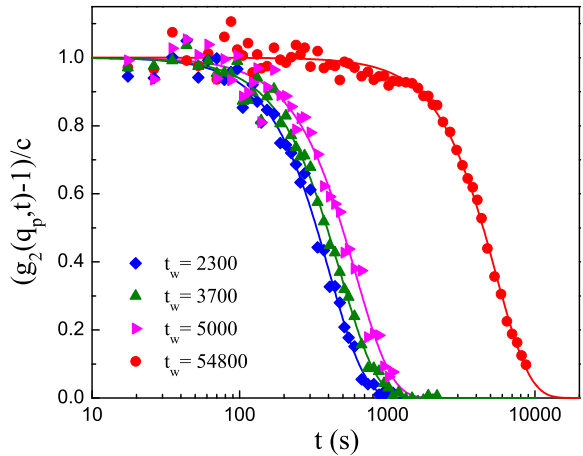}
\vspace{-0.8cm}
\caption[]{Selection of the intensity correlation functions measured with XPCS at $T=373$ K for different waiting times from temperature equilibration. On increasing the waiting time, the decay clearly shifts toward longer time scales due to physical aging. The data are reported together with the resulting fit obtained using equation \eqref{KWW}.}
\label{Fig2}
\vspace{-0.5cm}
\end{figure}
%%%%%%%%%%%%%%%%%%%%%%%%%%%%%%%%%%%%%%%%%%%%%%%%%%%%%%%%%%%%%%%%%%%%%%%%
The intensity correlation functions decorrelate completely on the experimental time scale, implying that the system is able to rearrange its structure on the probed spatial length scale of few \r{A}. The effect of aging is illustrated by the shift of the decay toward longer time scales upon increasing the sample age, in agreement with the analysis of the two times correlation functions reported in Figure \ref{Fig1}. As shown in Figure \ref{Fig2}, aging is so important that the decay time increases of almost a factor 10 in less than 13 hours. This fast aging is likely to be the reason of the well known rapid embrittlement of metallic glasses \cite{wu1990,murali2005}.\\
Interestingly, the correlation functions display a compressed, thus faster than exponential behavior, which can be well described by equation \eqref{KWW} with a shape parameter $\beta\sim 1.8$, independent of the sample age. Such a high value is in agreement with the results reported for other metallic glasses \cite{ruta2012,leitner2012}, while it cannot be explained within the current theories for the dynamics in the glassy state \cite{narayanaswamy1971,moyhinan1993,lubchenko2004}. Based on the observations of the dynamics in the liquid above the glass transition temperature, these theories predict a
stretched, thus with $\beta\leq1$, decay of the density fluctuations even in the glassy state. The idea is that the dynamic heterogeneities present in the liquid phase partially freeze below $T_g$, leading to a narrower distribution of relaxation times, and thus to an increase of the shape exponent, up to the limiting case of $\beta=1$. This explanation well describes the correlation functions calculated in numerical simulation \cite{kob1997} and the increase of $\beta$ measured in some polymeric glasses \cite{alegria1995,alegria1997}, but cannot take into account the behavior reported in Figure \ref{Fig2}. \\
From the comparison with other systems, it seems then clear that metallic glasses behave in a different universal way, independently of their atomic structure. In the case of Mg$_{65}$Cu$_{25}$Y$_{10}$, the unusual faster than exponential behavior has been attributed to the strain field generated by a random distribution of slowly-evolving sources of internal stresses stored in the system during the quench, and which can then be released upon annealing the system close to $T_g$ \cite{bouchaud2001,ruta2012}. This interpretation suggests the existence of a different kind of dynamics, rather than pure diffusion, in metallic glasses, which display many similarities with that of out-of-equilibrium soft materials, such as concentrated colloidal suspensions and nanoparticle probes in a glass former matrix \cite{ruta2012,cipelletti2000,bandyopadhyay2004,caronna2008,guo2009}. \\
In the case of Zr$_{67}$Ni$_{33}$ the shape parameter is even larger than the one reported for Mg$_{65}$Cu$_{25}$Y$_{10}$ ($\beta\sim1.5$ for a similar thermal path). This may suggest the presence of larger atomic mobility or more internal stresses in Zr$_{67}$Ni$_{33}$, a conclusion which however requires further investigations. While the value $\beta\sim1.5$ has been often observed in numerous experimental studies of soft materials and corresponds to the early-time regime proposed in the model of Bouchaud and Pitard \cite{bouchaud2001}, values of $\beta$ as large as those here reported for Zr$_{67}$Ni$_{33}$ have been observed only in another Zr-based metallic glass \cite{leitner2012} and in few other soft systems, and their interpretation is still far from being reached \cite{caronna2008,czakkel2011}.\\ 
%%%%%%%%%%%%%%%%%%%%%%%%%%%%%%%%%%%%%%%%%%%%%%%%%%%%%%%%%%%%%%%%%%%%%%%%
\begin{figure}
\includegraphics[width=9 cm]{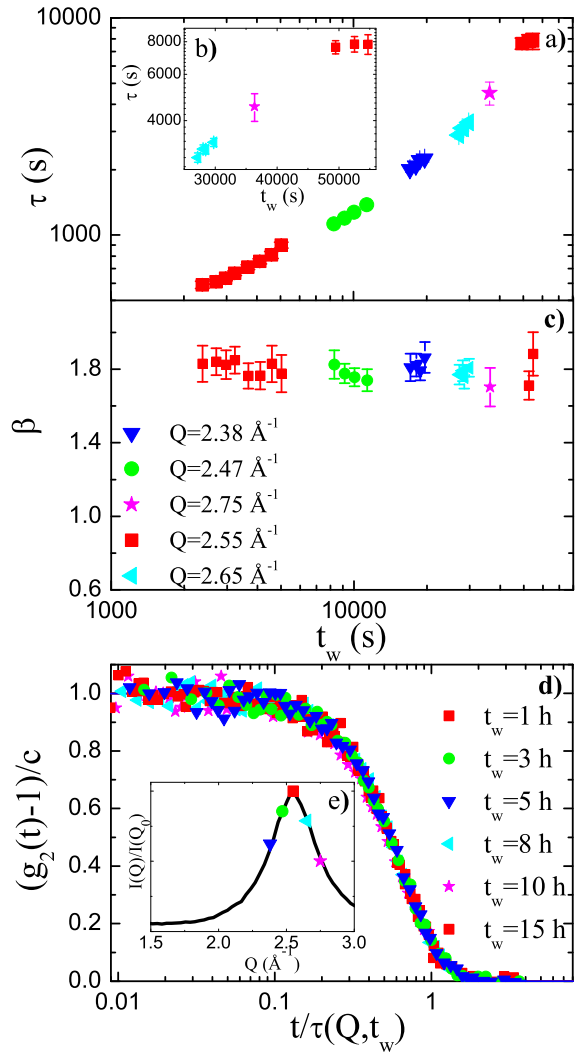}
\vspace{-0.8cm}
\caption[]{\textbf{a)} Waiting time dependence of the structural relaxation time measured at $T$=373 K for different wave-vectors around the first sharp diffraction peak ($q_p=2.55$ \r{A}$^{-1}$). The inset \textbf{b)} reports a zoom of the data taken at large waiting times. \textbf{c)} Corresponding shape parameter. \textbf{d)} Selection of the correlation curves corresponding to the data reported in \textbf{a)} as a function of the time rescaled by the structural relaxation time obtained from the analysis of the curves with the KWW model function of Eq. \eqref{KWW}. The inset \textbf{e)} indicates the position of the different investigated $q$s.}
\label{Fig4}
\vspace{-0.5cm}
\end{figure}
%%%%%%%%%%%%%%%%%%%%%%%%%%%%%%%%%%%%%%%%%%%%%%%%%%%%%%%%%%%%%%%%%%%%%%%%
The waiting time dependence of the structural relaxation times obtained from the analysis of the $g_2(q_p,t)$ is shown in Figure \ref{Fig4} a) (squares). Data corresponding to different $q$ values around the main peak of the diffraction pattern are also reported. As explained in the experimental details, the low scattered signal allowed us to measure just one $q$ vector for each detector setting. Consequently the different $q$s are taken at different $t_w$ from temperature equilibration. At all the investigated $q$s, $\tau$ grows rapidly with the sample age, in agreement with the results found in the Mg$_{65}$Cu$_{25}$Y$_{10}$ glass \cite{ruta2012}. Interestingly, the dynamics is characterized by structural rearrangements which require $\sim 10^3-10^4$ s thus, relatively short times with respect to the common idea of an almost frozen dynamics in the deep glassy state. These values are however in perfect agreement with those reported for both the Mg-based and Zr-based metallic glasses \cite{ruta2012,leitner2012} and suggest the presence of a peculiar dynamical behavior at the atomic length scale.
Hence, it would be interesting to investigate the nature of the atomic dynamics in the glassy state through the $q$ dependence of the structural relaxation time. Unfortunately, this information would require at each investigated $q$ a good knowledge of the aging in a wide $t_w$ range, which can be gained only by measuring the dynamics simultaneously for different wave-vectors. A higher incident flux, as the one available at free electron laser sources, will make this kind of studies possible.\\
Interestingly, the shape parameter does not display any dependence on $q$ or $t_w$, and remains constant with $\beta\sim1.8\pm0.08$ for all the investigated $q$s (Figure \ref{Fig4} c)). This means that it is possible to rescale all the data sets to a single master curve and define a time-waiting time-wave vector superposition principle in the glass. This scaling is reported in \ref{Fig4} d) where a selection of intensity auto correlation functions corresponding to the data reported in panel a) are plotted as a function of $t/\tau$. In the case of Mg$_{65}$Cu$_{25}$Y$_{10}$, $\beta$ was found to be temperature and $t_w$ independent in the fast aging regime \cite{ruta2012}. Our results indicate that this parameter is also not affected by the choice of the wave-vector, at least for values of $q$ close to that of the maximum of the static structure factor (see Figure \ref{Fig4} e)). \\
As shown in Figure \ref{Fig4} a) and b), for $q=q_p$ and large $t_w$, the fast aging regime seems to change to a more stationary or very slow one, as also suggested by the analysis of the two times correlation function reported in the bottom panel of Figure \ref{Fig1}. The presence of a complex hierarchy of aging regimes has been observed also in Mg$_{65}$Cu$_{25}$Y$_{10}$ but for temperatures very close to the glass transition one ($T/T_g=0.995$). The results here reported for $T/T_g=0.577$ show that this second aging regime can be found also in the deep glassy state, after a sufficiently long annealing time. In the case of Mg$_{65}$Cu$_{25}$Y$_{10}$, the second aging regime was accompanied by a decrease in the shape parameter. As shown in Figure \ref{Fig4} c), in the present case the lack of a full complete decay of $g_2(t)$ for the highest $t_w$ leads to a larger error in the determination of $\beta$, which can be determined only for $t_w\geq 50000$ s and remains however similar to the value obtained at lower $t_w$.\\
From the energetic point of view, the abrupt crossover to an almost stationary dynamics suggests the existence of a deep energy minimum in the potential energy landscape, where the system remains trapped for a very long time before eventually jumping into a different configuration \cite{goldstein1969,stillinger1995}. A similar plateau has been recently observed in the enthalpy recovery of glassy polymers and has been interpreted as a thermodynamically stable state \cite{boucher2011}. This explanation cannot be extended to the case of metallic glasses. In fact, the comparison between XPCS and low temperature equilibrium viscosity data for Mg$_{65}$Cu$_{25}$Y$_{10}$ suggests instead a gradual approach toward equilibrium \cite{ruta2012}, in agreement with the sub linear aging regime found in other polymeric glasses and colloids \cite{struik1978,alegria1997,cipelletti2005}.\\
Following the procedure reported in Ref. \cite{ruta2012}, we can describe the short $t_w$ behavior of $\tau_{q_p}(t_w)$ using the phenomenological expression
\begin{equation}
\tau(T,t_w)=\tau_0(T)\exp(t_w/\tau^{\ast})
\label{aging}
\end{equation}    
where $\tau_0(T)$ is the value of $\tau$ for $t_w=0$ at the temperature equilibration time, and $\tau^{\ast}$ is a fitting parameter which describes the rate growth and is independent of temperature \cite{ruta2012}. A fast exponential growth with $t_w$ is evident also at the other investigated $q$s. However, in that cases the lack of information on the behavior at shorter $t_w$ does not allow for a more quantitative description of the aging dependence for $\tau$.\\
\begin{figure}
\includegraphics[width=9 cm]{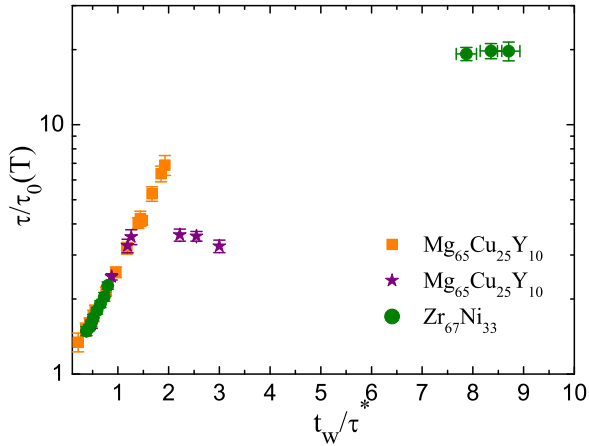}
\vspace{-0.8cm}
\caption[]{Structural relaxation times measured in two as quenched glasses of Mg$_{65}$Cu$_{25}$Y$_{10}$ at $T/T_g=0.995$ with different thermal histories and in Zr$_{67}$Ni$_{33}$ at $T/T_g=0.577$. All the data are normalized for the $\tau_0$ parameter obtained from the analysis of the fast aging with the empirical model function \eqref{aging} and are reported as a function of the time rescaled by the $\tau^{\ast}$ parameter. The data for Mg$_{65}$Cu$_{25}$Y$_{10}$ are taken from Ref. \cite{ruta2012}.}
\label{Fig5}
\vspace{-0.8cm}
\end{figure}
For $q=q_p$ we find $\tau_0=407\pm6$ s and $\tau^{\ast}=6490\pm170$ s. The latter value is quite close to the one reported in ref. \cite{ruta2012} for Mg$_{65}$Cu$_{25}$Y$_{10}$ at $T/T_g=0.995$ and it is of the same order of magnitude of that previously reported in the case of colloidal gels and Laponite \cite{cipelletti2000,schosseler2006}, thus strengthening the idea of a universal out-of-equilibrium behavior. The comparison with the Mg-based metallic glass suggests that the physical aging is very similar for both systems and that it should be possible to rescale the rapid growth of $\tau(t_w)$ to a unique curve, independently on the microscopic details of the system. This scaling is shown in Figure \ref{Fig5}, where the structural relaxation times of Zr$_{67}$Ni$_{33}$ at $T/T_g=0.577$ are compared to those measured in Mg$_{65}$Cu$_{25}$Y$_{10}$ at $T/T_g=0.995$ and for two different thermal paths. While squares and circles correspond to as-quenched samples heated up to a given temperature, stars refer to a state reached after having partially annealed the glass at higher temperatures without equilibration \cite{ruta2012}. All the data are rescaled for the corresponding $\tau_0$ parameter obtained from the analysis of the fast aging using equation \eqref{aging} and are reported as a function of the time rescaled by the $\tau^{\ast}$ parameter. For short $t_w$, all the data fall into a unique master curve, suggesting a common origin of the fast aging regime in metallic glasses. Differently, the crossover to the subsequent almost stationary dynamics strongly depends on the temperature and thermal path, and is reached only for very long annealing times (circles) or after particular thermal treatments (stars). 

\section{Conclusions}

In conclusion, we have studied the physical aging in a Zr$_{67}$Ni$_{33}$ metallic glass, by following the temporal evolution of the structural relaxation process at the interatomic length scale. We find that the decay of the density fluctuations can be well described by compressed, thus faster than exponential correlation functions, suggesting a peculiar origin for the atomic dynamics \cite{cipelletti2003,madsen2010}. This behavior seems to be universal in metallic glasses \cite{ruta2012,leitner2012}, and cannot be explained by the well-known theories for glasses dynamics, which predict a stretched exponential shape of the correlation functions, even in the glassy state \cite{narayanaswamy1971,moyhinan1993,lubchenko2004}. Model systems such as hard spheres and Lennard-Jones are unable to reproduce these intriguing dynamic effects \cite{kob1997,masri2010}. In these model systems, indeed, the dynamics appears to be always characterized by a broad spectrum of relaxations.
Conversely, very similar results have been previously reported for out of equilibrium soft materials, suggesting a universal microscopic dynamics for these very different systems \cite{ruta2012}. In these systems, the compressed behavior arises from ballistic-like motion due to the presence of internal stresses in the out of equilibrium state \cite{cipelletti2003}. It would be interesting to check whether this is also the case for metallic glasses. However, a $q$ dependence study would require a much stronger signal than the one achievable even at a third generation synchrotron source as the one used for the present study. %This will soon be possible thanks to the unique coherent properties of the X-ray beams available at Free Electron Laser sources. 
Notwithstanding, indication of the presence of distinct processes of atomic motion in metallic glasses have been reported for different Zr-based metallic glass-formers studied with nuclear magnetic resonance \cite{tang1999}. In these systems the diffusive, collective motion is found to be accompanied by hopping which becomes dominant in the glassy state, and could then be related to the observed compressed relaxation process in our glass.\\
Albeit we could not get information on the $q$ dependence of the relaxation time due to the impossibility of measuring simultaneously different $q$s, we have investigated the nature of the dynamics at different wave-vectors and sample ages. We find that the shape of the correlation functions does not show any dependence neither on the sample age nor on the wave-vector, at least for $q$s around the maximum of the static structure factor of the system, which is at $q_p=2.55$ \r{A}$^{-1}$. These findings allow then to rescale all the correlation curves on the top of each other and define a time-waiting time-wave-vector superposition principle in the fast aging regime.\\
The value of $\beta$ seems to depend on the details of the system, being larger in Zr-based systems than in the previously reported Mg$_{65}$Cu$_{25}$Y$_{10}$ \cite{ruta2012}. As both these glasses were obtained with the same thermal protocol, we speculate that the observed difference could be related to a larger atomic mobility in the first system and to the consequent presence of larger internal stresses. Despite this small difference, the comparison between the two metallic glasses suggests the existence of a universal dynamical behavior, also indicated by the scaling of the fast aging regime for the relaxation time to a single curve, independently of the microscopic details of the systems. Additional theoretical and experimental works is clearly required to elucidate this remarkable dynamical behavior of metallic glasses.\\
 \\
%%%%%%%%%%%%%%%%%%%%%%%%%%%%%%%%%%%%%%%%%%%%%%%%%%%%%%%%%%%%%%%%%%

We acknowledge F. Zontone, H. Vitoux, K. L'Hoste, L. Claustre for technical support during the XPCS experiments and E. Pineda an P. Bruna for providing the samples.

%\newpage %Just because of unusual number of tables stacked at end
%\bibliography{RefMgCuY}% Produces the bibliography via BibTeX.

\end{document}